\documentstyle[prl,aps,amssymb,psfig]{revtex}

\newcommand{\be}{\begin{equation}}
\newcommand{\ee}{\end{equation}}
\newcommand{\bea}{\begin{eqnarray}}
\newcommand{\eea}{\end{eqnarray}}
\newcommand{\eq}[1]{Eq.~(\ref{#1})}
\newcommand{\fig}[1]{Fig.~\ref{#1}}

\begin{document}

\preprint{\bf PREPRINT}
\columnsep0.1truecm
\draft

\twocolumn[\hsize\textwidth\columnwidth\hsize\csname@twocolumnfalse\endcsname

\title{Damping of Growth Oscillations}
\author{
  Harald Kallabis$^{(1,2,5)}$,
  Lothar Brendel$^{(1,2)}$,
  Pavel \v{S}milauer$^{(3)}$,
  Joachim Krug$^{(4)}$, and
  Dietrich E.~Wolf$^{(2,5)}$
}
 
\address{
(1)~H\"ochstleistungsrechenzentrum, Forschungszentrum J\"ulich,
52425 J\"ulich, Germany\\
(2)~FB 10, Theoretische Physik, Gerhard--Mercator--Universit\"at GH Duisburg,
47048~Duisburg, Germany\\
(3)~Institute of Physics, Czech Academy of Sciences, Cukrovarnick\'{a} 10, 162~53~Praha 6, Czech
Republic\\
(4)~Fachbereich Physik, Universit\"at GH Essen, D-45117 Essen, Germany\\
(5)~Center for Polymer Studies, Boston University, Boston, MA 02215,
USA
}

\maketitle

\begin{center}
\today
\end{center}

\begin{abstract}

Computer simulations and scaling theory are used to investigate the
damping of oscillations during epitaxial growth on high-symmetry
surfaces.  The crossover from smooth to rough growth takes place after
the deposition of $(D/F)^{\delta}$ monolayers, where $D$ and $F$ are
the surface diffusion constant and the deposition rate, respectively,
and the exponent $\delta=2/3$ on a two-dimensional surface. At the
transition, layer-by-layer growth becomes desynchronized on distances
larger than a layer coherence length proportional to $\ell^2$, where
$\ell$ is a typical distance between two-dimensional islands in the
submonolayer region of growth.

\end{abstract}

]

\narrowtext

\section{1 Introduction}
Layer-by-layer or Frank-van der Merwe growth \cite{bauer} is a growth
mode observed in molecular beam epitaxy and other deposition 
methods which allows precise control of chemical composition of layers 
down to atomic thickness.  It is therefore particularly well suited for 
the fabrication of novel electronic devices.

The key microscopic processes in layer-by-layer growth are deposition
of atoms onto a high-symmetry surface and diffusion of adatoms on
the surface.
The adatoms meet and form dimers which then grow into islands of 
monoatomic height whose edges capture most of the adatoms during 
the deposition of one monolayer. When the island edges become
less available
due to coalescence of islands, formation of dimers and 
islands in the next layer begins.  The density of atomic steps -- and
all other quantities sensitive to the surface morphology -- thus
oscillates in time. 

Generically, these oscillations are damped: Layer-by-layer growth is
only a transient.  Possible reasons \cite{wolf97} include (i) cessation 
of periodic formation of islands on the surface and transition to step 
flow growth \cite{orr,broehl}, or (ii) roughening of the surface
\cite{evans}.  If the substrate temperature is increased, damping 
becomes stronger in the first and weaker in the second case, allowing 
to discriminate between the two.  Ignoring the possibility of 
inhomogeneous deposition (cf. Ref.~\cite{comsa}), surface roughening
can have two different sources.  If interlayer
transport is inhibited by step-edge barriers, one obtains the growth
instability predicted by Villain \cite{villain91}.  If no such
instability occurs, the surface may still roughen due to fluctuations
in the intensity of the deposition rate. Only the latter case
is considered in this paper.

The question of how long layer-by-layer growth persists is of immediate 
practical importance. If the answer is known, one can devise and optimize
annealing schedules for growing thicker films while maintaining a
smooth surface. For this purpose, it is important to know both the
damping time $\tilde t$ {\em and\/} the length scale over which
layer-by-layer growth is synchronized. This {\em layer coherence length\/}
$\tilde \ell$ is a new characteristic length that
determines, e.g., the annealing time needed to reestablish 
a flat surface \cite{annealing} before growth can be continued.

A theory has been proposed recently \cite{Damping,MRJK} which predicts
that the damping time and the layer coherence length depend on the
typical distance $\ell$ between islands in the submonolayer region
of growth as

\be
F\tilde t \propto \ell^{4d/(4-d)} \quad \mbox{and} \quad
\tilde \ell \propto \ell ^{4/(4-d)}.
\label{result}
\ee

In this paper, we present the theory of oscillations damping
and detailed numerical evidence of the 
validity of its predictions based on extensive computer simulations 
of a minimal, {\em one-parameter\/} model at surface 
dimension $d=2$. The methods of extracting the damping time 
and the layer coherence length from the surface morphology
evolution are outlined and thoroughly discussed.

The characteristic length $\ell$ (and thus also $\tilde t$ and 
$\tilde\ell$) has a power-law dependence on the ratio $D/F$ of 
the surface diffusion constant to the deposition rate:

\be
\ell \propto (D/F)^{\gamma}
\ee
(see \cite{zinsmeister} and references given in \cite{wolf97,kkw}).
The exponent $\gamma$ depends on the dimensionality $d$ of the surface
and the (possibly non-integer) dimension $d_f$ of the islands. It also
depends on whether or not desorption of adatoms or diffusion of dimers
or larger clusters is negligible.  Finally, $\gamma$ is a function of
the critical cluster size $i^*$ for the formation of a stable
nucleus. For the case considered here in more detail ($d=2$, $d_f=2$,
$i^*=1$, no desorption, immobile clusters), the value is $\gamma =
1/6$.

The layer coherence length $\tilde \ell$ as well as the damping time
$\tilde t$ play the roles of natural cutoffs in the continuum growth
equation at small length and time scales.  For $t \gtrsim \tilde t$
one expects that the surface exhibits self--affine scaling
\cite{family_vicsek}:

\be
w(t) \simeq a_{\perp}(\xi(t)/\tilde \ell \ )^{\zeta} 
\qquad {\mbox{and}} \qquad
\xi(t) \simeq \tilde \ell \ (t/\tilde t \ )^{1/z}.
\label{w_scaling}
\ee
Here $w$ is the root mean square variation of the film thickness (the
surface width), $a_{\perp}$ the thickness of one atomic layer (which,
for convenience, is set to one in the following), and $\xi$ the
correlation length up to which the surface roughness has fully
developed at time $t$. $\zeta$ is the roughness exponent and $z$
the dynamical exponent.  The dependence of $\tilde\ell$ and $\tilde t$
on the microscopic growth parameters will be derived next.

\section{2 Theoretical results}
Coarse-graining the surface configuration at a given time over a
length scale of the order of $\ell$, one can write down an evolution
equation for the variable $h(x,t)$.  Since particle desorption can be
neglected under conditions typical for molecular beam epitaxy, the 
equation can be written in the form of a conservation law,

\be
\partial_t h(x,t) = - \nabla j(x,t) + \eta(x,t).
\label{evolution}
\ee
$j$ is the surface diffusion current, and $\eta$ is 
white noise with second moment

\be
\langle\eta(x,t)\eta(y,s)\rangle = F
\delta^d(x-y)\delta(t-s),
\ee
which describes the fluctuations  in the deposition rate.
It was proposed by Villain \cite{villain91} that in growth processes
far from equilibrium where local chemical potentials along the
surface are ill defined, diffusion currents should be driven by
gradients in the growth-induced, nonequilibrium adatom density $n$
\cite{equicont},

\begin{equation}
{\bf j}= - D \nabla n.
\label{villain_current}
\end{equation}
On a singular surface, the balance between deposition and capture of
adatoms at steps leads to a stationary adatom density $n = n_0$ of the
order of \cite{villain92} $n_0 \simeq (F/D) \ell^2$.  On a vicinal
surface, the adatom density is reduced due to the presence of additional
steps.  However, this effect is felt only if the miscut $m=|\nabla h|$
exceeds $1/\ell$, in which case $n \simeq (F/D) m^{-2}$.  In terms of a
coarse-grained description of the surface this implies that the local
adatom density depends on the local miscut or surface tilt. A useful
interpolation formula which connects the regimes $m \ll 1/\ell$ and $m
\gg 1/\ell$ is \cite{politi}

\bea
\label{n(m)}
n(\nabla h) &=& \frac{n_0}{1 + (\ell \nabla h)^2} \\
&\simeq &(F/D) \ell^2 - 
(F/D) \ell^4 (\nabla h)^2 + \dots
\eea
Inserting the leading quadratic term of this gradient expansion into
(\ref{villain_current}), which is appropriate for describing
long-wavelength fluctuations around the singular orientation, one obtains

\be
j = \nabla \lambda (\nabla h)^2
\label{current}
\ee
with

\begin{equation}
\lambda = F \ell^4.
\label{villaineq}
\end{equation}

Considering \eq{evolution} and \eq{current} one sees that the physical
dimension of $\lambda$ is (length)$^4$/(time$\cdot$height).  Within the
continuum description, the only characteristic length and time scales
are the layer coherence length and the damping time, whereas the
lattice constant $a_{\perp}$ has been chosen as a unit of height.
Therefore

\be
\lambda \propto \tilde\ell^4 / \tilde t
\label{dimanal}
\ee
on dimensional grounds.

Finally, the number of particles deposited during the time $\tilde t$
onto an area $\tilde \ell^d$ is $F \tilde t \ \tilde \ell^d \pm{}(F
\tilde t \ \tilde \ell^d )^{1/2}$. Thus the fluctuation of the film
thickness over the distance $\tilde \ell$ is $w(\tilde t) \simeq
\sqrt{F \tilde t \ \tilde \ell^d}/\tilde \ell^d$.  At $\tilde t$ this
should be the thickness of about one atomic layer, $w(\tilde t) \simeq
1$, which results in

\begin{equation}
F \tilde t = \tilde \ell^d.
\label{F3}
\end{equation}

Combining (\ref{villaineq}), (\ref{dimanal}) and (\ref{F3}) one
obtains Eq.~(\ref{result}), or

\be
F \tilde t \simeq (D/F)^{\delta}
\qquad {\mbox{and}} \qquad
\tilde \ell \simeq (D/F)^{\delta/d}
\label{result1}
\ee
with the exponent 

\be
\delta = \frac{4d}{4-d}\,\gamma.
\label{delta}
\ee
Notice that the layer coherence length $\tilde \ell$ is substantially
larger than the characteristic distance $\ell$ between islands 
($\tilde \ell \simeq \ell^2$ at $d=2$).

\section{3 Model} 
In our model, atoms are deposited onto the (100) surface of a simple
cubic lattice with the rate of $F$ atoms per unit time and area.  The
surface size is $L\times L= 128\times 128$ \cite{FS}.  Atoms with no
lateral neighbors are allowed to diffuse with diffusion constant
$D$. Atoms with lateral neighbors are assumed to be immobile so that,
e.g., dimers are immobile and stable.  Growth commences on a flat
substrate, $h(x,0)=0$ for all sites $x$.  On deposition at $x$,
$h(x,t)$ is increased by one.  We neglect barriers to interlayer
transport (Ehrlich--Schwoebel barriers \cite{ESB})  so that the only
parameter of the model is the ratio $D/F$.

\begin{figure}[htb]
\centerline{\psfig{figure=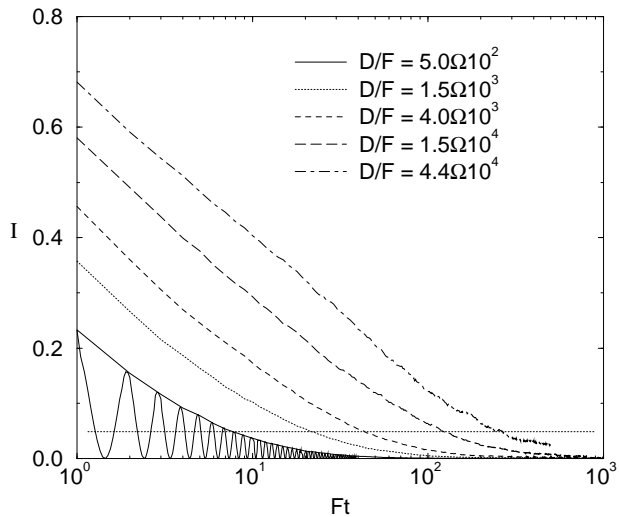,height=7cm,angle=270}}
\caption{
  Semilogarithmic plot of the upper envelopes of the oscillations of
  the kinematic intensity $I$ as a function of time. The minima of the
  oscillations are zero, as indicated schematically for $D/F=500$.
  Five different $D/F$ values are evaluated,
  spanning almost two decades.  The dotted line denotes the value
  $I=0.05$ used for the determination of the damping time.
}
\label{I.fig}
\end{figure}

\section{4 Damping time}  
First we present the results for the damping time extracted from
kinematic intensity data,

\be
I \equiv \langle (N_{\rm even}-N_{\rm odd})^2/L^2 \rangle
\ee 
(see \fig{I.fig}; $N_{\rm even}$ $(N_{\rm odd})$ denotes the number of
atoms in even (odd) layers). The brackets $\langle\dots\rangle$ denote
averaging over different runs.  The same analysis was done for the
surface width with equivalent results for the damping time.

\begin{figure}[htb]
\centerline{\psfig{figure=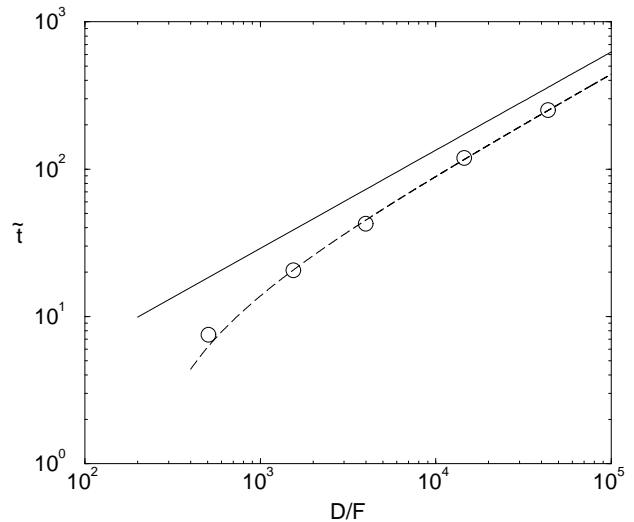,height=7cm,angle=270}}
\caption{
  Damping time $\tilde t$ determined from 
  Fig.~\protect\ref{I.fig}, as a function of
  $D/F$.  The straight line has a slope 2/3, in agreement with the
  theoretical prediction. The dotted line is the best fit according to
  \protect\eq{newscaling}.
}
\label{DTimeFrac}
\end{figure}

The kinematic intensity oscillates between zero and maxima which
decrease until the oscillations vanish.  We measure $\tilde t$ as the
time where the maxima of the kinematic intensity drop below $I=0.05$
\cite{remark}. The results are shown in Fig.~\ref{DTimeFrac}.

Obviously there are strong corrections to scaling which can be
attributed to an offset $\tilde t_0>0$:

\be
\label{newscaling}
\tilde t = A_{\tilde t}\left(\frac{D}{F}\right)^\delta
- \tilde t_0.
\ee
$\tilde t_0$ plays the role of a cutoff for the validity of our
scaling theory.  $(D/F)_0\equiv(\tilde t_0/A_{\tilde t})^{1/\delta}$
can be interpreted as that value of $D/F$ below which the oscillations
are not observable anymore \cite{ModAnsatz}.

A three--parameter fit to the data shown in Fig.~\ref{DTimeFrac}
gives an exponent

\be
\delta = 0.69 \pm 0.05,
\label{eq:res}
\ee
where the error bar reflects the variations obtained when the evaluation
method is modified \cite{remark,ModAnsatz} or when the data for the
surface width are evaluated in the same way.  This value is in good
agreement with the theoretical prediction of $\delta=2/3$ (see
\eq{delta}) for compact islands.

\begin{figure}[htb]
\centerline{\psfig{figure=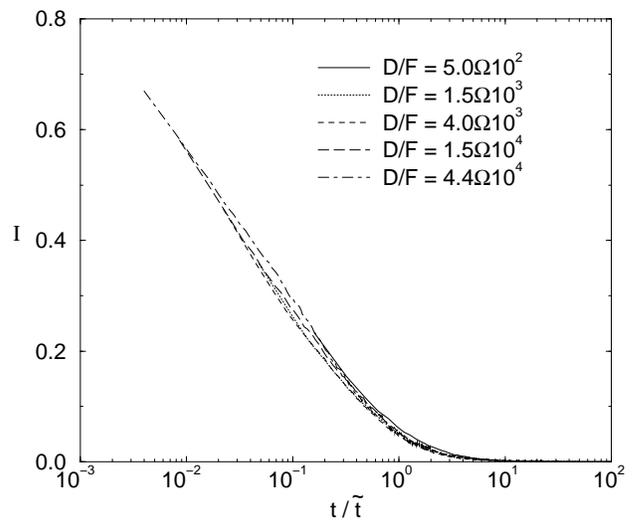,height=7cm,angle=270}}
\caption{
  Scaling the deposition time with the damping time
  [\protect\eq{newscaling}] leads to a good data collapse of the curves
  presented in Fig.~\protect\ref{I.fig}.
}
\label{Iscal.fig}
\end{figure}

In our simulation, neither detachment of adatoms from islands nor edge
diffusion are considered. Therefore, the islands are fractal for large
diffusion lengths \cite{PimpinelliFractal,AmarFamily,Jensen} with the
fractal dimension $d_f\simeq 1.72$ of two--dimensional diffusion
limited aggregation \cite{DLA}.  Then $\gamma$ changes from
$\gamma=1/6$ to $\gamma\simeq 0.175$.  This leads to theoretical
prediction $\delta\simeq 0.70$ which is also within the error bars of
Eq.~(\ref{eq:res}) \cite{Details}.  However, in the present case the
values of $D/F$ are sufficiently small, so that this complication may
be ignored.

The scaling plot of the kinematic intensity with the time divided by the
damping time according to \eq{newscaling} (see Fig.~\ref{Iscal.fig})
confirms the validity of the approach used. 

\section{5 Layer coherence length}
\begin{figure}[htb]
\centerline{\psfig{figure=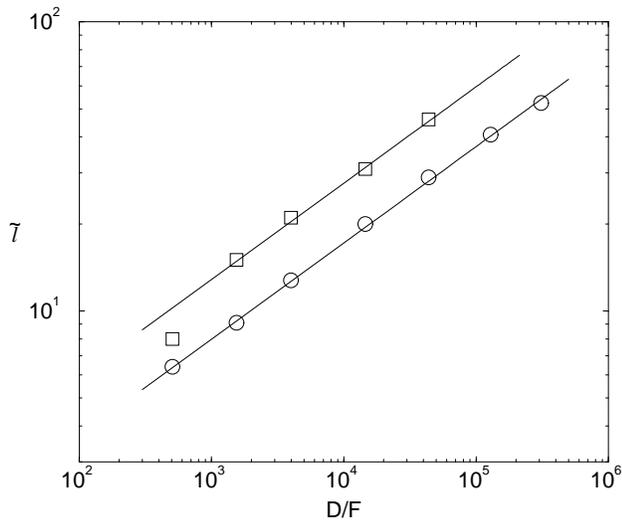,height=7.0cm,angle=270}}
\caption{
  $\tilde \ell$ as a function of $D/F$, measured using the height
  difference correlation function (squares), and finite-size analysis
  (circles), respectively.  The solid lines have the theoretically
  predicted slope 1/3 for compact islands, see \protect\eq{result1}.
}
\label{ltilDdF.fig}
\end{figure}

The measurement of the height difference correlation function
\cite{Details}

\be
G(x,t) \equiv \langle [h(x_0,t)-h(x_0+x,t)]^2 \rangle,
\ee
evaluated at $t=\tilde t$ shows that $G$ has a maximum. This can be
explained as follows.  
At $t=\tilde t$ the probability to find the surface at the same height
as at a reference point $x_0$ is minimal at a distance $x-x_0$
corresponding to the layer coherence length. For larger distances,
deviations from the average height are essentially uncorrelated. At
very small distances, their correlation is positive, while around
$\tilde \ell$ they are anticorrelated.  This is how the data denoted
by the squares in \fig{ltilDdF.fig} were obtained.  The result is in
good agreement with the predicted exponent,
cf. Eq.~(\ref{result1}). (Note that this method could be also used for
experimental determination of $\tilde\ell$.)

An alternative method of measuring $\tilde\ell$ is to carry out a
finite-size analysis in the following way.  The surface does not
roughen when the linear system size $L$ is smaller than $\tilde
\ell$. In this case, the amplitude of the growth oscillations becomes
stationary after a transient time, and the oscillations never die out.
We monitored the variance

\be
A^2(t)\equiv\langle w^2(t)\rangle_{[t,t+\tau]} 
- \langle w(t)\rangle^2_{[t,t+\tau]}
\ee
of the surface width during the layer completion time $\tau\equiv 1/F$,
where $\langle \dots \rangle_{[t,t+\tau]}$ denotes the time average
over the interval $[t,t+\tau]$.  Its stationary value decreases with
increasing system size and ultimately becomes equal to the statistical
fluctuations of $w$ when the system size is big enough so that the
oscillations can die out completely.  The values of $\tilde\ell$,
denoted by the circles in \fig{ltilDdF.fig} represent the linear
system size $L$ at which the stationary value of $A(t)$ drops below
$0.37$. Both methods of measuring $\tilde\ell$ are in excellent
agreement with each other and with the theoretical prediction
\cite{Details}.

\section{6 Conclusions and outlook}
We have presented a theory for the
damping of growth oscillations caused by
kinetic roughening. We have shown that the results of numerical
simulations of a minimal model compare very favorably with the theory,
and directly determined two key quantities, the damping time and
the layer coherence length.
The instability associated with barriers to interlayer transport
\cite{villain91} may compete with the kinetic roughening mechanism as
a source of oscillation damping \cite{DocBrendel}. This, as well as the
transition to step--flow growth on vicinal surfaces will lead to
different power laws for the damping time and the layer coherence
length. This remains for future research.

The results of this paper can be directly verified by diffraction or
real-space surface sensitive techniques. The determination of the
damping time and, in particular, of the layer coherence length as a
function of growth conditions should be possible using the methods
outlined above.

\section{Acknowledgements}
Useful conversations with Martin Rost are gratefully acknowleged.
D.~E.~W.~ acknowledges support by DFG within SFB 166 {\em Strukturelle
und magnetische Phasen\"uberg\"ange in \"Ubergangsmetall-Legierungen
und Verbindungen\/}.  
J.~K.\ acknowledges support by DFG within SFB 237 {\em Unordnung und
grosse Fluktuationen}.
P.~\v{S}. acknowledges the financial support of Alexander von Humboldt
Foundation and Volkswagen Stiftung.
H.~K.\ acknowledges support by the German Academic
Exchange Service within the Hochschulsonderprogramm III.


\begin{references}
\bibitem{FS} 
  At high $D/F$ ratios, simulations were performed using systems of
  linear size $L=200$ to ensure that finite size effects are
  negligible: Within the statistical fluctuations there was no
  observable change in the measured quantities.
\bibitem{remark}
  In order to check the robustness of our results, we repeated the
  same analysis for $I=0.1$ and obtained the same exponent $\delta$
  within the error bar.
\bibitem{ModAnsatz}
  We checked that another possible Ansatz $\tilde t=A_{\tilde
  t}[D/F-{\mbox{const.}}]^\delta$ only changes the exponent within the
  error bar.
\end{references}
\end{document}